\documentclass[preprint,authoryear,12pt]{elsarticle}
\usepackage{amssymb}

\journal{}

\begin{document}

\begin{frontmatter}

\title{Search strategies for Trojan asteroids in the inner Solar System}

\author[label1]{M. Todd\corref{cor1}}
\ead{michael.todd@icrar.org}

\author[label2]{D. M. Coward}

\author[label1]{M. G. Zadnik}

\address[label1]{Department of Imaging and Applied Physics, Bldg 301, Curtin University, Kent St, Bentley, WA 6102, Australia}
\address[label2]{School of Physics, M013, The University of Western Australia, 35 Stirling Hwy, Crawley, WA 6009, Australia}

\cortext[cor1]{Corresponding author.}

\begin{abstract}
Trojan asteroids are minor planets that share the orbit of a planet about the Sun and librate around the L4 or L5 Lagrangian points of stability. They are important because they carry information on early Solar System formation, when collisions between bodies were more frequent. Discovery and study of terrestrial planet Trojans will help constrain models for the distribution of bodies and interactions in the inner Solar System. 

We present models that constrain optimal search areas, and strategies for survey telescopes to maximize the probability of detecting inner planet Trojans. We also consider implications for detection with respect to the Gaia satellite, and limitations of Gaia's observing geometry.
\end{abstract}


\end{frontmatter}

\section{Introduction}
Trojan asteroids are minor planets that share the orbit of a planet about the Sun, and librate around the Lagrangian points of stability
that lie $60^{\circ}$ ahead of (L4), or behind (L5), the planet in its orbit. Trojans represent the solution to Lagrange's famous triangular problem and appear to be stable on long time-scales (100 Myr to 4.5 Gyr) \citep{pil99,sch05} in the N-body case of the Solar System. This raises the question whether the Trojans formed with the planets from the Solar nebula or were captured in the Lagrangian regions by gravitational effects. Study of the Trojans therefore provides insight into the early evolution of the Solar System.

About 5000 Jupiter Trojans are currently known to exist. Some searches have been conducted for Earth Trojans (ET) \citep{dun83,whi98,con00}. A search covering approximately 0.35 deg$^{2}$ by \citet{whi98} resulted in a crude upper population limit of fewer than three objects per square degree, down to $R=22.8$. A subsequent search by \citet{con00} covering about nine square degrees, down to $R=22$, failed to discover any ETs. Recent examination of data from the WISE satellite has resulted in the discovery of the first known ET \citep{con11}. Among the other terrestrial planets, four Trojans have been discovered in the orbit of Mars. Current models \citep{mik90,tab00a} suggest that this represents less than a tenth of the Mars Trojan population.

This paper describes probability distributions synthesized from existing models. This allows us to constrain optimal search areas and examine strategies to maximize the probability of detecting Trojans. We also examine the possibility of detection of Trojans by the Gaia satellite. 

\section{Models}
\subsection{Earth Trojans}

\begin{figure}
\includegraphics{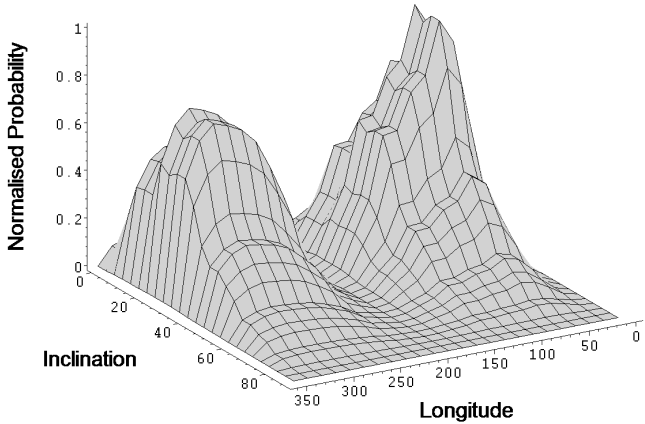}
\caption{\label{fig:figure1}Normalised probability contour for Earth Trojan bodies by Inclination and Heliocentric Longitude (degrees). The figure shows peak detection probabilities for longitudes consistent with the classical Lagrangian points but that bodies, while co-orbital with Earth, are unlikely to be co-planar.}
\end{figure}

A synthesis of a stable orbit inclination model \citep{mor02} and heliocentric longitude model \citep{tab00b} was used to identify probability regions for existence of bodies (Fig. \ref{fig:figure1}). With limits established for the regions of interest the sky area in the heliocentric frame can be easily determined using the standard solid angle integral $\int\!\!\!\int_{S} \left(\mathbf{r} \cdot \mathbf{n}\right)/r^{3} dS$ (where $\mathbf{r}$ is the radius vector, $\mathbf{n}$ is the unit normal vector, and $r = \left| \mathbf{r} \right|$). Calculation of the geocentric solid angle, necessary for Earth-based observations, requires a transformation from the heliocentric reference. A numerical integration is performed with which we can calculate the sky area presented by any region from any position. This enables determination of the sky area for an Earth-based observer, or a space-based instrument such as the Gaia satellite which will be positioned at Earth's L2 Lagrangian point.

\begin{figure}
\includegraphics{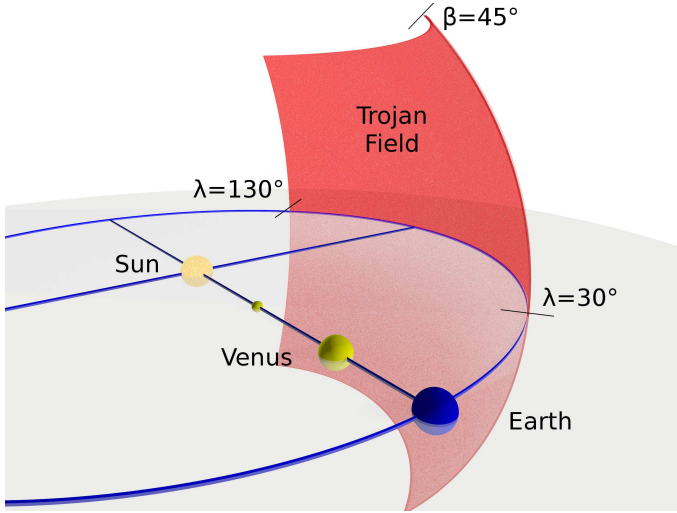}
\caption{\label{fig:figure2}Perspective illustration of Earth
Trojan (L4) field. The field is bounded by Heliocentric longitude limits
$30^{\circ} \le \lambda \le 130^{\circ}$ and inclination limit $\beta \le 45^{\circ}$. A complementary
field exists in the trailing Lagrangian L5 region.}
\end{figure}

The ET fields (Fig. \ref{fig:figure2}) are bounded by the upper inclination limit (FWHM) of $\sim45^{\circ}$ \citep{mor02} and heliocentric longitude limits (FWHM) of $30^{\circ}\lesssim\lambda\lesssim130^{\circ}$ (L4 region) and $240^{\circ}\lesssim\lambda\lesssim340^{\circ}$ (L5 region) \citep{tab00b}. The heliocentric solid angle of each of these regions is 2.468~sr (8100 deg$^{2}$). The geocentric solid angle%
\footnote{The Python code used to calculate these areas is available
from the first author on request.%
} of the L4 ET field is 1.066~sr (3500 deg$^{2}$). The time required to completely survey this area, several hours even for widefield survey telescopes (see Table \ref{tab:Comparison-of-survey}), is greater than the amount of time the field is visible per day.

\begin{table*}
\caption{Comparison of survey telescopes showing the required time
to survey the entire field.\label{tab:Comparison-of-survey}}

\begin{tabular}{llcccccc}
\hline 
Telescope  & Limiting  & Exposure  & FOV  & \multicolumn{2}{c}{Entire field } & Instrument capabilities \tabularnewline
 & mag. &  &  & (FOVs) & (Time) &  \tabularnewline
\hline 
Catalina  & $V\sim20$ & 30~s & 8.0~deg2 & 437 & 3.6h & \citep{dra09} \tabularnewline
Pan-STARRS & $R\sim24$ & 30~s & 7.0~deg2 & 499 & 4.2h & \citep{jed07} \tabularnewline
LSST$^{\dagger}$  & $r\sim24.7$ & 30~s & 9.6~deg2 & 364 & 3.0h & \citep{jon09} \tabularnewline
Gaia$^{\ddagger}$ & $V\sim20$ &  & 0.69~deg & 63 &  & \citep{mig07} \tabularnewline
\hline 
\end{tabular}

$^{\dagger}$The Large Synoptic Survey Telescope (LSST) is still in the development phase (www.lsst.org).

$^{\ddagger}$Gaia will operate in a continuous scanning mode where the CCD array will be read out at a rate corresponding to the angular
rotation rate of the satellite (6h period). The FOV value represents the number of rotations by Gaia. Gaia's specific precession parameters are not considered so values should be considered as representative.
\end{table*}

\begin{figure}
\includegraphics{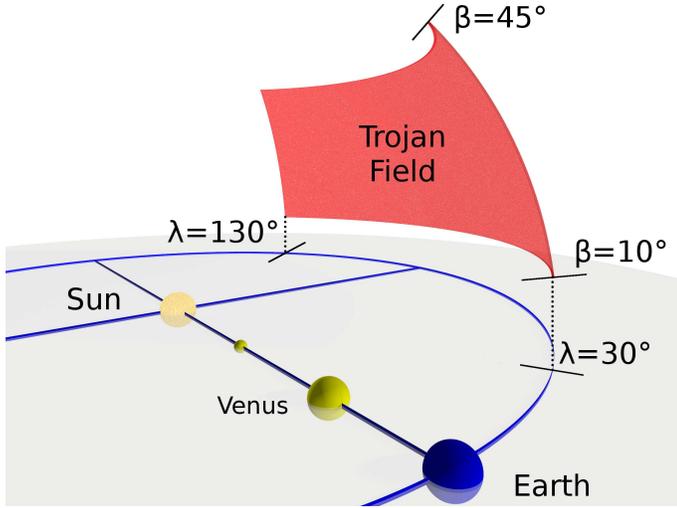}
\caption{\label{fig:figure3}The Earth
Trojan (L4) target field ranges from Heliocentric longitude
$30^{\circ} \le \lambda \le 130^{\circ}$ and latitude $10^{\circ} \le \beta \le 45^{\circ}$. A complementary field exists in the trailing Lagrangian L5 region. This illustration represents the search field in which a body will be observable some time during its orbit.}
\end{figure}

Selecting inclination limits of $10^{\circ}\lesssim\beta\lesssim45^{\circ}$ (Fig. \ref{fig:figure3}), based on the FWHM of the binned inclination distribution in \citet{mor02}, reduces the sky area of the field to about 1300 deg$^{2}$ and includes $\sim74$ per cent of those simulated bodies. A widefield telescope can survey this field in a single session of about 1.5 hours (Table \ref{tab:Comparison-of-survey2}). A second set of observations is normally necessary for moving object detection. These paired sessions should be repeated at intervals of no more than three months since the field spans 100$^{\circ}$ longitude. Such a programme will take one year to complete. This strategy requires a significant amount of telescope time on these nights. This is not considered an optimal strategy as one aim is to have minimal impact on other activities.

A strategy which further reduces the per-session time is to survey a region bounded by the longitude limits and lower inclination limits to detect ETs as they cross the ecliptic. This reduces the sky area of the region to 900 deg$^{2}$ and results in a corresponding decrease in telescope time per-session (Table \ref{tab:Comparison-of-survey2}). The duration of this programme is six months as any Trojans detected in that six-month period would be crossing either to the North or to the South. However the observations must be repeated at more frequent intervals since, in this region, the ET will have a higher apparent motion than at the highest or lowest points in its orbit, as indicated in \citet{mor02}. As a consequence it will cross this region relatively quickly. The result is that the total time requirement for the programme is not significantly reduced.

Gaia will operate in a continuous scanning mode and survey the whole sky, by rotating and precessing at a predetermined rate, to a limiting magnitude of $V \sim 20$ many times during its mission \citep{mig07}. Without considering the precession rate, or its effect on Gaia's coverage of the Trojan regions, we reduce the calculation to consider only the minimum possible number of rotations to survey the entire field (Table \ref{tab:Comparison-of-survey}). In this context the FOVs and time are incompatible to comparison because the operation is unalterable. Also, Gaia will effectively survey each field twice per cycle, taking narrow, slightly overlapping, swaths of the sky with each rotation.

Examination of Gaia's mode of operation leads us to the concept of surveying a swath and using Earth's revolution about the Sun to sweep out the field described in Figure \ref{fig:figure3}. This greatly reduces the time requirement per session as indicated in Table \ref{tab:Comparison-of-survey2}. A swath $10^{\circ}$ wide reduces the per-session sky area to about 140 deg$^{2}$, reducing the imaging time to a few minutes. This strategy requires paired observing sessions on a weekly basis. The field is redefined at monthly intervals to repeat the survey of the target field. The length of this programme is one year, however this programme has the advantage of minimal time per session.. These observations can be conducted at the end of twilight, with the primary science missions continuing as normal during the night. We consider this strategy could be employed for an extended period with minimal impact on regular activities.

\begin{table*}
\caption{Comparison of survey telescopes showing the required time for different survey strategies.\label{tab:Comparison-of-survey2}}

\begin{tabular}{lcccccccc}
\hline 
Telescope  & \multicolumn{2}{c}{Entire field } & \multicolumn{2}{c}{Restricted field} & \multicolumn{2}{c}{Ecliptic region} & \multicolumn{2}{c}{$10^{\circ}$ swath} \tabularnewline
 & (FOVs) & (Time) & (FOVs) & (Time) & (FOVs) & (Time) & (FOVs) & (Time) \tabularnewline
\hline 
Catalina  & 437 & 3.6h & 163 & 1.4h & 112 & 56m & 18 & 9m \tabularnewline
Pan-STARRS & 499 & 4.2h & 186 & 1.6h & 128 & 64m & 20 & 10m \tabularnewline
LSST & 364 & 3.0h & 136 & 1.2h & 94 & 47m & 15 & 8m \tabularnewline
\hline 
\end{tabular}

\end{table*}

\subsection{Mars Trojans}

\begin{figure}
\includegraphics{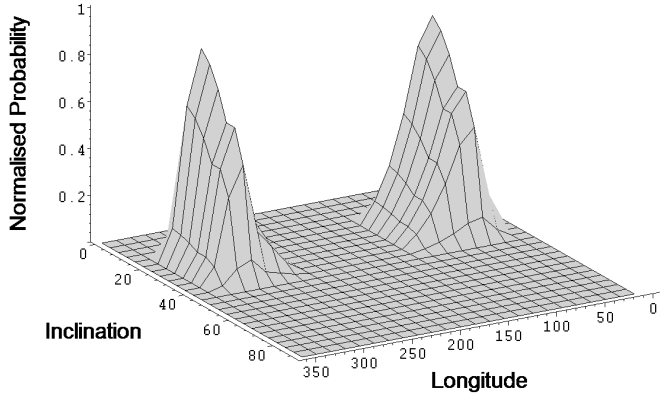}
\caption{\label{fig:figure4}Normalised probability
contour for Mars Trojan bodies by Inclination and Heliocentric Longitude
(degrees). The figure shows peak detection probabilities for longitudes
consistent with the classical Lagrangian points but that bodies, while
co-orbital with Mars, are unlikely to be co-planar.}
\end{figure}

\begin{figure}
\includegraphics{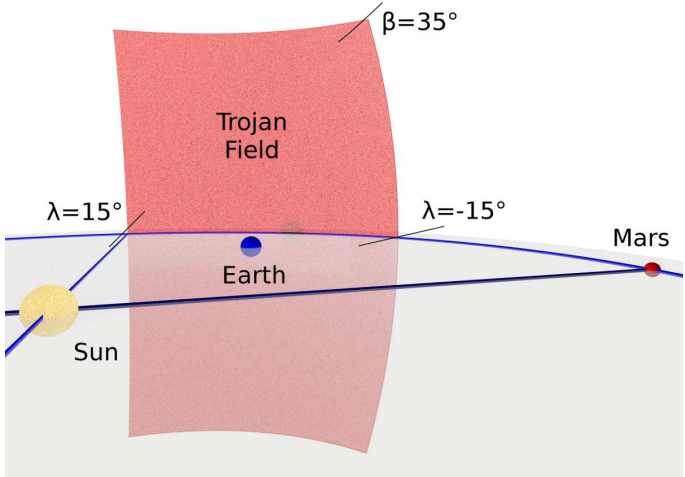}
\caption{\label{fig:figure5}
Mars Trojan (L4) target field is bounded by Heliocentric longitudes
$50^{\circ} \le \lambda \le 80^{\circ}$ and inclination limit $\beta \le 45^{\circ}$. A complementary
field exists in the trailing Lagrangian L5 region.}
\end{figure}

Similar problems exist in planning observations to search for additional Mars Trojans (MT). The MT fields are bounded by the upper inclination limit (FWHM) of $35^{\circ}$ \citep{sch05} and heliocentric longitude limits (FWHM) of $50^{\circ} \lesssim \lambda \lesssim 80^{\circ}$ and $290^{\circ} \lesssim \lambda \lesssim 315^{\circ}$ \citep{tab00b} for the L4 and L5 regions respectively. A synthesis of these models identifies probability regions for existence of bodies (Fig. \ref{fig:figure4}). These regions each define a sky area (Fig. \ref{fig:figure5}) of about 9500 deg$^{\circ}$ at opposition, almost three times larger than ET field. The changing geometry between Earth and Mars complicates a comprehensive analysis. At this time we examine the strategy considered optimal for the ET search, that of taking swaths of the field and using Earth's revolution about the Sun to sweep out the entire region, in the case of the MT field at opposition. A detailed study will be the subject of future work in this area.

We find that, at opposition, it is still most efficient to take swaths of the MT field as Earth revolves past the field in its orbit. The sky area surveyed per-day can be the same as that for the ET field. The geocentric latitudes are greater than for the ET field, but the longitude range can be reduced accordingly. The major difference to be noted is that the range in Right Ascension and Declination can be redefined for each pair of observations to survey regions at opposition. Before and after opposition, when the relative position of Earth results in a smaller sky area for the MT fields, other strategies for surveying this region could be employed. This will be the subject of future investigation.

\subsection{Telescope surveys}

For ground-based telescopes, geographic location is a consideration. Observing with a telescope in a particular hemisphere increases the observing window in that hemisphere. This is at the expense of the complementary field, i.e. Northern Hemisphere sites have fields at Northern latitudes visible for a greater duration at the expense of the amount of time for which a Southern latitude field is visible due to Earth's axial inclination. Gaia, at the L2 Lagrangian point, will spend equal amounts of time surveying both Northern and Southern latitude fields.

\begin{figure}
\includegraphics{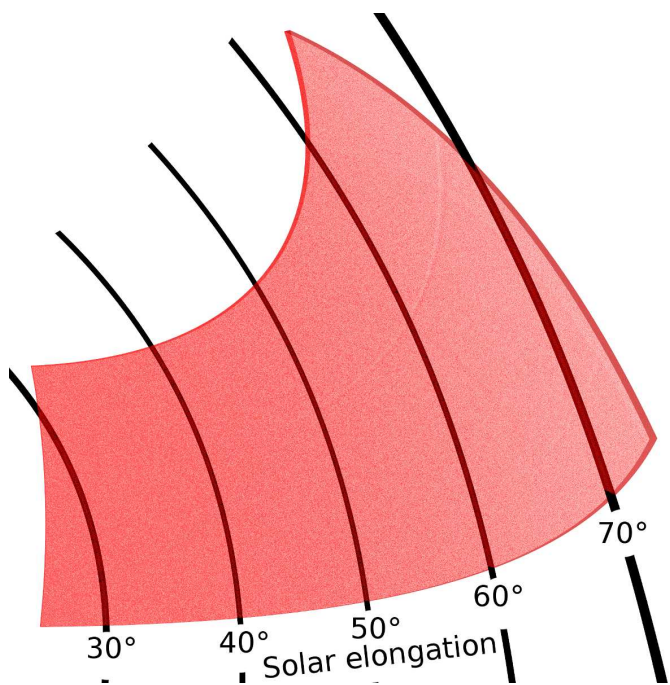}
\caption{\label{fig:figure6}Solar elongation across the Earth Trojan field ranges from $25^{\circ}$ to $75^{\circ}$. This illustration represents the field from a geocentric perspective.}
\end{figure}

The limitation in observable longitudes must also be considered. If observing begins when the Sun's altitude is $20^{\circ}$ below the local horizon, a minimum observing altitude of $20^{\circ}$ will survey longitudes with a Solar elongation of $40^{\circ}$. This still permits the greater part of the ET field, at greater elongations (Fig. \ref{fig:figure6}), to be surveyed although low altitudes will be subject to greater atmospheric extinction. 

Despite the small FOV of the Gaia satellite it has been included for comparison. Gaia does not suffer the limitation of local horizon and airmass affecting the available observing time. However, this advantage is mitigated by its limiting magnitude of $V=20$. Gaia has a defined orbit which prevents it observing targets of opportunity, however its primary mission is to perform an all-sky survey. A similar observing restriction exists for Gaia where, by design, it is limited to a minimum Solar elongation of $45^{\circ}$. This corresponds to an elongation of $\sim 48^{\circ}$ for a ground-based telescope.

Observations of ETs are time-limited by the Earth-Sun-object geometry. This geometry also causes them to appear to be slow-moving but repeat imaging can be performed on subsequent days. MTs are less problematic for the obvious reason that observing opportunities do not have the same time constraints that are imposed by the specific geometry for ET observations. 

While this delay between follow-up images introduces other variations due to such things as change in atmospheric conditions, seeing etc
this could be compensated for by some image convolution. Some telescopes are implementing image processing systems designed specifically for asteroid detection (e.g. Pan-STARRS+MOPS -- Moving Object Processing System) where the asteroid is an apparently stationary transient because it is a very distant slow moving object \citep{jed07}. This method could be applied to nearer objects which are apparently slow moving as a result of the particular Earth-Sun-object positions at the time of observation.

\section{Results}

Surveys of the entire ET field within the chosen limits are impractical. The sky area of each field is too great to cover in the available time with current survey telescopes. However it is possible to survey a region restricted by the upper and lower inclination limits.  These surveys can be conducted at the beginning (L5 region) and end (L4 region) of the night. However, while possible, it may be impractical as this occupies a significant amount of telescope time on those nights. As the field spans $\sim 90^{\circ}$ in longitude the intervals between surveys would nominally be three months. This bears the risk of missing low-inclination ETs completely through not having yet entered the field or of having just exited the field. It also bears the risk of detecting ETs shortly before exiting the field towards the ecliptic, possibly requiring follow-up observations in a less favourable orientation to the local horizon. Reducing the intervals to two months results in a small amount of oversampling, reducing the risk of missing an ET. Assuming that only the Northern or Southern field could be surveyed depending on the location of the telescope, the programme length would be 1 year, the orbital period of an ET. 

It is possible to conduct a survey in the ecliptic plane between latitude 0 and the lower limit of the field to search for ETs as they cross the ecliptic. This reduces the necessary time to survey the region as the ecliptic region is smaller. The time saving per session is about 30 per cent. In addition the length of the programme is reduced to six months, as any Trojans detected in that period would be crossing either to the North or to the South. In this region the ET will have a higher apparent motion, crossing the region relatively quickly, and more frequent sampling is required. This outweighs the benefit of the shorter programme as the total time requirement is not significantly different to that required to survey the entire field over the period of one year. 

Attempting to survey the entire ET field in a single session is challenging. However, observing a swath of sky in a particular range of Right Ascension and Declination and using Earth's revolution about the Sun to survey the field as it crosses the observed region requires minimal time each session, i.e. two sessions per week for one year. The observed region is redefined at monthly intervals so that the field is imaged again. This results in some oversampling of the field but has the benefit that lost nights due to adverse weather conditions become less critical to the overall programme.

Surveys of the MT field are only possible when the Sun-Earth-Mars geometry permits observations. A detailed study will be the subject of future work. In our initial assessment we find that surveys of the MT field at opposition are, in many ways, more difficult than the ET field as it occupies a much larger sky area. However, the field can be surveyed using the method of observing a swath of sky. As Earth passes the field during its revolution about the Sun the observations will progressively image the entire field. The range in Right Ascension and Declination must be redefined for each pair of observations so that the imaged region progresses across the MT field with Earth's revolution. 

We note that Gaia's precession cycle, which limits observations to Solar elongations $>~45^{\circ}$ \citep{mig07}, also prevents observations at Solar elongations $>~135^{\circ}$. This restricts Gaia from observing at opposition, hence such observations can only be made by other telescopes. The possibility of Gaia detecting MTs within its observable region will be addressed in future work.

\section{Summary and Future Work}

Despite the thousands of known Jupiter Trojans very few inner planet Trojans have been discovered. Simulations \citep{mor02} have predicted the existence of a number of Earth Trojans, but currently only one is known to exist \citep{con11}. We note that this object, 2010~TK$_7$, would pass through the region described in this paper during its orbit. The prospect of detecting additional ETs is limited by the small amount of time available each day due to the Earth - Sun - ET geometry. 

This paper has identified the region of highest probability for detection of ETs. We suggest an optimal strategy of observing a sub-region of the ET field and using Earth's revolution about the Sun to progressively survey the field. This approach takes only a few minutes per session, two days per week, and is readily achievable by a survey telescope such as Catalina or Pan-STARRS. While this method requires a programme of continued observations for one year, the total time commitment for the programme is a few tens of hours spread throughout that year.

A similar observing strategy is also suggested for the Mars Trojan field during its passage through opposition. The different, and changing, geometry of Earth with respect to MTs makes a detailed study more involved than the relatively static geometric condition of the ETs. Mars is our nearest planetary neighbour with known Trojans, with many more MTs predicted to exist \citep{tab99}. The peculiarities of the changing geometry and the implications for a search for additional Mars Trojans will be explored in greater depth in the future.

The specific observing geometry of the Gaia satellite and its position at Earth's L2 Lagrangian point will be examined in more detail in future work. Gaia will not share the limitations of ground-based telescopes of atmosphere and local horizon. The primary limitations identified for Gaia are the limiting magnitude $(V \sim 20)$ and its precession cycle which restricts observations to Solar elongations between $45^{\circ}$ to $135^{\circ}$ \citep{mig07}.  Results of detailed simulations for detection of Earth Trojans, with particular regard to the detection limits and observational mode of operation of Gaia, will be reported in the future.

Follow-up after any initial detection by additional observations or other actions may be necessary to enable orbit computations and confirmations of discovery. Some of these tasks may be best handled by telescopes other than the survey telescope which made the initial detection. In the case of Gaia, a number of ground-based telescopes as a follow-up network for Solar System object detection is being developed \citep{thu10}.

\section*{Acknowledgments}
The authors would like to thank the anonymous referees whose comments and suggestions significantly improved the final version of this manuscript. M. Todd thanks the organisers of the Gaia workshop (Pisa 2011) for providing a fertile environment for discussing Gaia science. D.M. Coward is supported by an Australian Research Council Future Fellowship.

\bibliographystyle{elsarticle-harv}

\end{document}